\documentclass[a4paper,fleqn,usenatbib]{mnras}

\usepackage{newtxtext,newtxmath}

\usepackage[T1]{fontenc}
\usepackage{ae,aecompl}

\usepackage{graphicx}	

\usepackage{amsmath}	
\usepackage{amssymb}	

\usepackage{units}

\usepackage[usenames,dvipsnames]{color}

\usepackage{multicol}
\usepackage{multirow}

\newcommand{\nuclei}[2]{\ensuremath{\mathrm{^{#1}#2}}}

\newcommand{\hydrogen}[1][1]{\nuclei{#1}{H}}
\newcommand{\helium}[1][4]{\nuclei{#1}{He}}

\newcommand{\carbon}[1][12]{\nuclei{#1}{C}}
\newcommand{\nitrogen}[1][14]{\nuclei{#1}{N}}

\newcommand{\code}[1]{\textsc{#1}}
\newcommand{\mesa}{\code{MESA}}
\newcommand{\MESA}{\mesa}

\newcommand{\Msun}{\ensuremath{\mathrm{M}_{\sun}}} 
\newcommand{\Msunyr}{\ensuremath{\rm \Msun\,yr^{-1}}} 

\newcommand{\Rsun}{\ensuremath{\mathrm{R}_{\sun}}}

\newcommand{\qH}{\ensuremath{q_{{\rm H}}}}
\newcommand{\XH}{\ensuremath{X_{{\rm H}}}}

\title[Hot subdwarfs from He WD mergers]{Hot subdwarfs formed from the merger of two He white dwarfs}
\author[Schwab]{
Josiah Schwab$^{1}$\thanks{Hubble Fellow; E-mail: jwschwab@ucsc.edu}
\\
$^1${Department of Astronomy and Astrophysics, University of California, Santa Cruz, CA 95064, USA} \\
}

\pubyear{2018}

\begin{document}
\label{firstpage}
\pagerange{\pageref{firstpage}--\pageref{lastpage}}
\maketitle

\begin{abstract}
  We perform stellar evolution calculations of the remnant of the
  merger of two He white dwarfs (WDs).  Our initial conditions are
  taken from hydrodynamic simulations of double WD mergers and the
  viscous disc phase that follows.  We evolve these objects from
  shortly after the merger into their core He-burning phase, when they
  appear as hot subdwarf stars.  We use our models to quantify the
  amount of H that survives the merger, finding that it is difficult
  for $\ga \unit[10^{-4}]{\Msun}$ of H to survive, with even less
  being concentrated in the surface layers of the object.  We also
  study the rotational evolution of these merger remnants.  We find
  that mass loss over the $\sim \unit[10^4]{yr}$ following the merger
  can significantly reduce the angular momentum of these objects.  As
  hot subdwarfs, our models have moderate surface rotation velocities
  of $30-100\; {\rm km\,s^{-1}}$.  The properties of our models are
  not representative of many apparently-isolated hot subdwarfs,
  suggesting that those objects may form via other channels or that
  our modelling is incomplete.  However, a sub-population of hot
  subdwarfs are moderate-to-rapid rotators and/or have He-rich
  atmospheres.  Our models help to connect the observed properties of
  these objects to their progenitor systems.
\end{abstract}



\section{Introduction}

The hot subdwarf stars (of spectral types O and B) are He-burning
stars with H envelopes too small ($\la \unit[10^{-2}]{\Msun}$) to
support H shell-burning; see \citet{Heber2009} and \citet{Heber2016a}
for recent reviews of these objects.  (Hereafter, we use sdOB as a
generic shorthand for hot subdwarf stars, rather than as an indicator
for any particular sub-population.)  Many sdOB stars are found in
binary systems, where interaction has previously stripped the
H-envelope.  However, some sdOB stars are observed to be presently
single.  As described by \citet{Webbink1984}, the merger of two He
white dwarfs (WDs) provides a natural mechanism for producing these
objects.  In such a scenario, most of the H was ejected during the
earlier binary evolution that formed the tight double He WD binary and
then the merger itself causes the ignition of He burning.  After a
phase of adjustment, the object settles into a core He-burning (CHeB)
configuration.

\citet{Iben1986b} discuss this scenario in more detail and
qualitatively consider several properties of the merged objects.  They
suggest that the H layers initially present on the He WDs will be
buried by the turbulent mixing that occurs during the merger.  Much of
the H may be consumed, though diffusion may return some to the surface.
They also discuss the rotation of these objects and speculate that
even if a significant fraction of the pre-merger orbital angular
momentum is retained by the remnant, this could be extracted by magnetic stellar winds
such that the final object might not be rapidly rotating.


\citet{Iben1990} constructs quantitative models of He WD mergers,
treating the merger as a super-Eddington accretion event.  These
models ignite He off-center and this He burning migrates inwards
through a series of flashes.  It reaches the center having burned only
a small fraction of the total He and the object settles into a CHeB
phase.  \citet{Saio1998} construct similar models using lower,
sub-Eddington accretion rates and find similar results.  Together,
these studies confirm the basic idea of merger-initiated He ignition.

\citet{Saio2000} further demonstrate that stellar models
during the phase where He-burning is migrating inwards are in
agreement with the observed properties of some extreme He stars.
Their interpretation is that objects such as V652 Her are the result
of He WD mergers and that they will subsequently evolve to be sdOB
stars.  Population synthesis studies of the formation of sdOB stars
indicate that the He WD merger channel can indeed play a significant
role in the formation of single sdOB stars \citep{Han2002, Han2003,
  Han2008}.

Recent work had improved the modelling of double He WD merger remnants and
more quantitatively compares the model properties to observations.
\citet{Zhang2012a} build models that approximate the merger as a mix
of fast and slow accretion.  Their models have surface temperatures,
gravities, and abundances (in particular N and C) in agreement with
the observed population of He-sdO stars.  \citet{Hall2016} address the
important question of how much H can survive the merger.  They show
that models with their estimated surviving H masses are in agreement
with the atmospheric properties of the observed population of
apparently-isolated, H-rich sdB stars.

Most studies of the remnants of double WD mergers (including the
aforementioned ones) circumvent the complicated hydrodynamic processes
of the WD merger and its aftermath by treating this as an accretion
phase.  Such models are relatively simple to construct and provide
valuable insights.  However, these approaches can make it difficult to
address questions related to the unmodeled material in the ``accretion
reservoir'', such as the evolution of its chemical composition or
angular momentum.

Numerical hydrodynamics has long been used to study double WDs that
undergo dynamically unstable mass transfer \citep[e.g.,][]{Benz1990}.
Such calculations show that the system reaches a configuration with the relatively
undisturbed primary WD surrounded by the tidally-disrupted secondary
WD.  The latter consists of an envelope of shock-heated material and a
rotationally-supported disc.

\citet{Shen2012} show that the remnant disc, which is unstable to the
magneto-rotational instability, evolves viscously on a timescale of
$\sim$hours \citep[see also][]{vanKerkwijk2010}.  \citet{Schwab2012} follow the evolution of these viscous
discs using multidimensional hydrodynamics calculations that include
an appropriate equation of state and an $\alpha$-viscosity.
Together, these papers show that the remnants evolve towards a spherical
end-state, where the rotationally-supported disc has been converted
into a hot, thermally-supported envelope.

This understanding suggests the limitations of modelling the long-term
evolution as the accretion of the low mass WD onto the massive WD.
Instead, the post-merger evolution is a stellar evolution problem
driven by the internal redistribution of heat and momentum and not by
the external accretion of mass.  Motivated by that understanding, this
work takes initial conditions from \citet{Schwab2012} and maps these
into the \MESA\ stellar evolution code.  We then evolve the models
forward to the CHeB phase (when they become sdOB-like) and
characterize their properties.

There are two primary questions that we seek to address with these
models: (1) how much hydrogen can survive the merger and later reach
the surface and (2) how much angular momentum can be shed as the
remnants evolve.  Already \citet{Iben1986b} discuss both of these
questions qualitatively; decades later, we seek to address them more
quantitatively.

In Section~\ref{sec:models} we describe the initial conditions for our
models and in Section~\ref{sec:schematic} show how they evolve towards
a CHeB phase and beyond.  In Section~\ref{sec:hydrogen} we address the question
of how much H can survive this evolution; in
Section~\ref{sec:rotation} we discuss the rotation of these objects.
In Section~\ref{sec:conclusions}, we summarize and conclude.

\section{Initial Conditions and Modelling Assumptions}
\label{sec:models}

We use release 10108 of \mesa\ \citep[Modules for Experiments in
Stellar Astrophysics;][]{Paxton2011, Paxton2013, Paxton2015,
  Paxton2018} to evolve our stellar models.  The input files necessary
to reproduce our results will be made available at
\url{http://mesastar.org}.  For detailed descriptions of specific
options, consult the \mesa\ website at
\url{http://mesa.sourceforge.net} and the \MESA\ instrument papers.

\subsection{Initial Conditions}

Our fiducial model is the result of an 0.2 \Msun + 0.3 \Msun WD merger
(model M05).  We also consider a 0.3 \Msun + 0.4 \Msun WD merger model
(model M07).  The smoothed particle hydrodynamics (SPH) calculations
of the merger are from \citet{Dan2011} and the post-merger viscous
evolution of these systems was performed in \citet{Schwab2012}.

Following a similar procedure to \citet{Schwab2016}, we reduce the
output of the multidimensional viscous evolution simulations to 1D
profiles.
The viscous evolution was performed with a grid-based code in
spherical coordinates assuming azimuthal symmetry.  Thus, we first
calculate the total mass, internal energy, and angular momentum of the
zones in each spherical shell.
Then, we use these values and the equation of state
\citep[the HELM EOS,][]{Timmes2000b} to produce profiles of the average density,
temperature and specific angular momentum corresponding to each spherical radius.
A similar
procedure also gives composition profiles; however, these are
essentially pure He because the H layers that would be present on the
He WDs were not included in the SPH simulations of the WD mergers and
little He burning took place during the merger and subsequent viscous
evolution.

\begin{figure}
  \centering
  \includegraphics[width=\columnwidth]{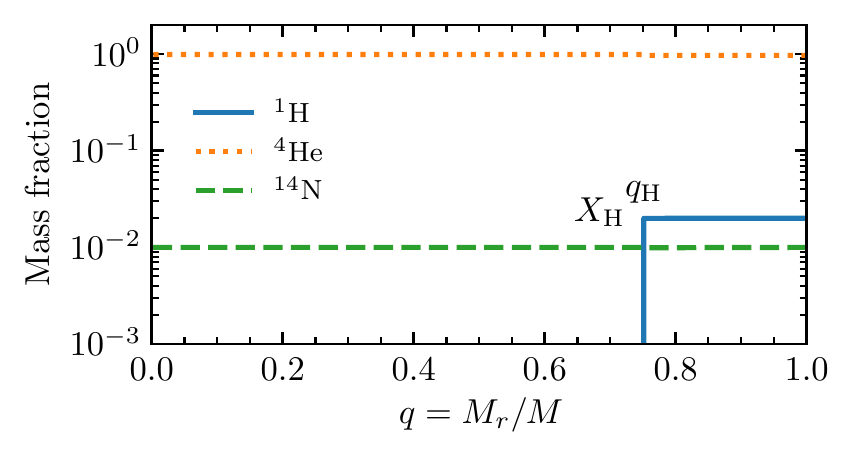}
  \caption{Parameterized composition used in the initial \MESA\
    models.  We include a mass fraction \XH\ of \hydrogen[1] in the
    outer layers (above normalized mass coordinate \qH) and a constant
    mass fraction of 0.01 \nitrogen[14] throughout.  The remaining
    material is \helium[4].}
  \label{fig:zp6-X}
\end{figure}

Since we are interested in exploring the fate of the H, we must make
an ad hoc assumption about where the H is located after the merger.
We initialize our models with a uniform \hydrogen[1] mass fraction
\XH\ in the outer layers (those above a normalized mass coordinate
\qH).  We use this to make the assumption that turbulent mixing during
the merger and its aftermath uniformly mixed the initial H throughout
the tidally disrupted secondary WD and perhaps destroyed H in the
inner regions.  In Section~\ref{sec:hydrogen} we will motivate and
discuss our choices of \XH\ and \qH\ and also demonstrate that our
results are not sensitive to their precise values.  We also assume the
material is roughly solar metallicity and put a uniform abundance of
1 per cent by mass of \nitrogen[14].  This is of particular
importance, since the H will primarily be destroyed via the CNO cycle.
Figure~\ref{fig:zp6-X} illustrates this schematic composition profile.
We use the \MESA\ default 8 isotope nuclear network \texttt{basic.net}
which covers H and He burning.

\begin{figure}
  \centering
  \includegraphics[width=\columnwidth]{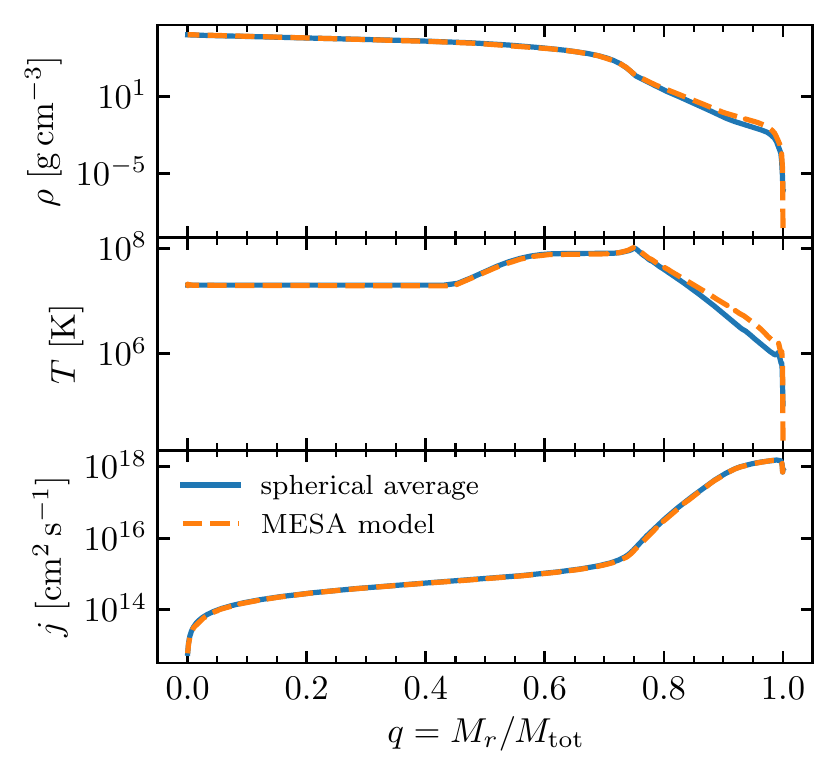}
  \caption{Comparison of density, temperature, and specific angular
    momentum profiles between spherical averages taken at the end of
    the hydrodynamic calculations of the viscous disc phase and the
    initial \MESA\ model for our fiducial model M05.}
  \label{fig:zp6-1D}
\end{figure}

We use the built-in \MESA\ \texttt{relax\_initial} controls to
construct a \MESA\ model with the given density, temperature and
composition profiles.
These relaxation routines construct \MESA\ models that match given
input profiles by evolving an arbitrary initial model with an extra
heating term, extra torque term, or extra composition change term.
These extra terms are chosen such that they guide the model towards
the desired structure on a relaxation timescale that is short compared
to other characteristic timescales in the model.  This proceeds for a
large number of relaxation timescales such that the structure of the
\MESA\ model becomes close to the input profiles.  (A exact match may
not be possible given that the input profiles may not satisfy the
equations solved by \MESA.)
Figure~\ref{fig:zp6-1D} shows that the input
spherical average profiles are approximately realized in the \MESA\ model.  Some
differences do appear in the outer layers ($q\ga0.75$) reflecting
departures from spherical symmetry and hydrostatic equilibrium in
these regions.  We then load these models into \MESA\ and evolve them
forward in time.

\subsection{Modelling Assumptions}

Our fiducial model does not include element diffusion in order to
reduce its computational cost.  However, we do run models including
element diffusion.  By virtue of using a recent \MESA\ release, this
takes advantage of improvements of the treatment of diffusion in
degenerate regions \citep[see section 3 in][]{Paxton2018}.

Beginning from the initial angular momentum profile given by the
viscous evolution calculations, our models continue to approximately
include the effects of rotation.  \MESA\ treats rotation in the
shellular approximation \citep[see section 6 in][]{Paxton2013}.  The
angular momentum transport and rotationally-induced mixing is
performed in a diffusive framework as in \citet{Heger2000a}.  We
include the processes of Eddington-Sweet circulation, the
Goldreich-Schubert-Fricke instability, the Solberg-H{\o}iland
instability, the secular shear instability, and the Spruit-Tayler
dynamo.  The rotational corrections to the stellar structure equations
are included via the factors $f_{\rm T}$ and $f_{\rm P}$
\citep{Endal1976}.  Note that these values have enforced minimum
limits that are triggered if the outer layers approach critical
rotation.  As we will show in Section~\ref{sec:rotation}, material
often reaches or exceeds critical rotation.  The results from 1D
calculations are necessarily particularly uncertain in this regime.
Nonetheless, this represents a first effort towards quantitatively
characterizing the role of rotation.

As is well-known, the duration of the CHeB phase depends on the mixing
processes that set the size of the convective core
\citep[e.g.,][]{Salaris2017}.  The work of \citet{Schindler2015}
explores many of the important choices related to this issue when
making \MESA\ models of sdB stars.  We use the OPAL C- and O-enhanced
(``Type II'') opacities \citep{Iglesias1996}.  We adopt mixing
parameters from \citet{Paxton2018} that make use of the improved
treatment of convective boundaries described in that work. We do not
include additional processes such as convective overshoot that might
further extend the core.  We do not attempt to tune the mixing
parameters to match asteroseismic constraints on the convective core size
\citep{VanGrootel2010a, VanGrootel2010b, Charpinet2011}, as that is
beyond the scope of this work.

\section{Schematic Evolution}
\label{sec:schematic}

\begin{figure*}
  \centering
  \includegraphics[]{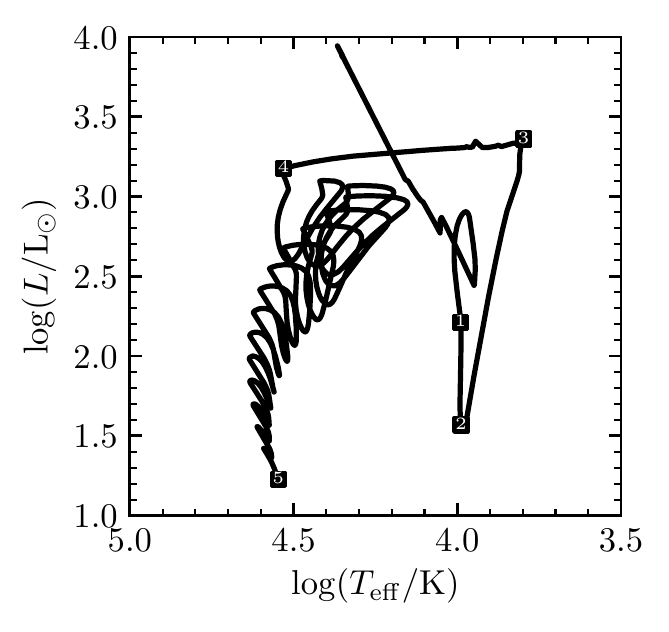}
  \includegraphics[]{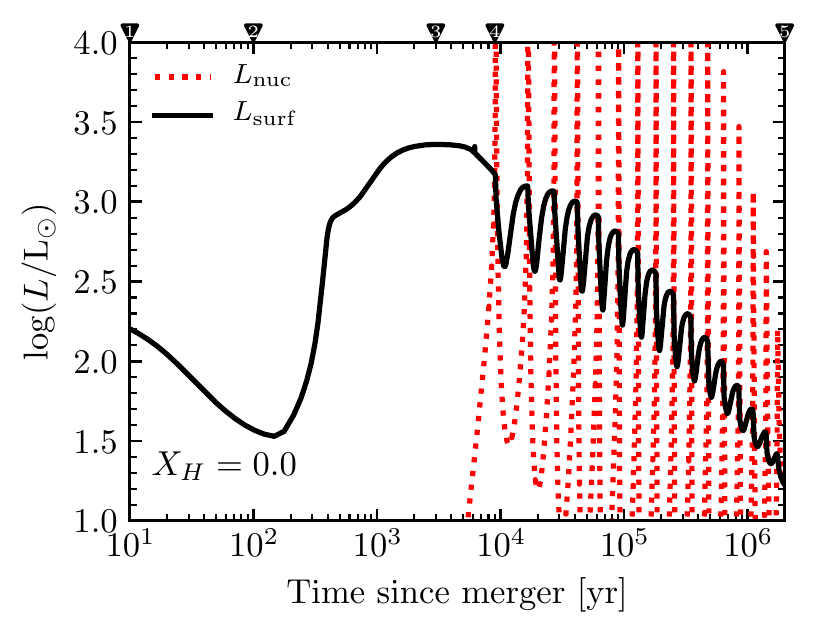}
  \caption{Evolution of the M05 model up to the onset of the CHeB
    phase.  The left panel shows the HR diagram.  The right panel
    shows the luminosity as a function of time.  There, the solid line
    shows the surface luminosity; the dashed line shows the nuclear
    luminosity.  This model has no H, so the nuclear energy release is
    from He burning.  The shared y-axis gives a correspondence between
    the two panels; as an aid, numbered squares are placed on the left
    track at the times marked by numbered triangles at the top of the
    right panel.}
  \label{fig:Ls}
\end{figure*}

Before we discuss our models in more detail, we briefly demonstrate
their evolution from the immediate post-merger state to the CHeB
phase.  Figure~\ref{fig:Ls} shows this evolution for the fiducial model M05.
The signature of the first $\sim\unit[10^2]{yr}$ of the evolution is
driven by the initial properties of the outer layers.  These are the
least spherical, most rotationally supported regions at the end of the
viscous calculations.  Thus, they are the most uncertain regions in
our \MESA\ modelling (see also Figure 10 and surrounding discussion).
We do not focus on the details of this early evolution.

\begin{figure}
  \centering
  \includegraphics[width=\columnwidth]{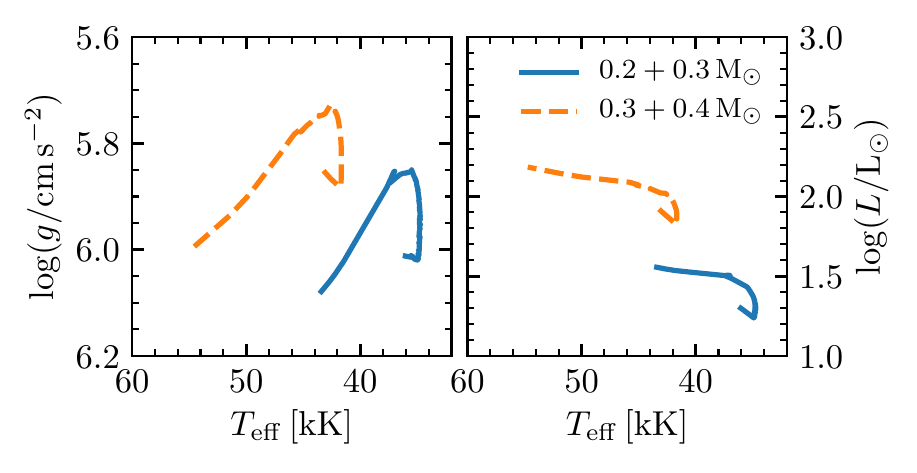}
  \caption{Location of the models on the Kiel diagram (left) and HR
    diagram (right) during their He core burning phase.  The models
    shown have essentially no H envelope.}
  \label{fig:gT-and-LT}
\end{figure}

Beginning around $\sim\unit[10^2]{yr}$ and continuing for
$\sim\unit[10^3]{yr}$, the remnant expands as it thermally adjusts.
This is followed by a $\sim\unit[10^4]{yr}$ phase in which the remnant
radiates away much of the thermal energy deposited during the
merger.\footnote{To clearly demonstrate that the energy radiated is
  that deposited in the merger, the M05 model we show in
  Figure~\ref{fig:Ls} has no H, and so has negligible nuclear energy
  release before $\unit[10^4]{yr}$.}  As their thermal support is
lost, the non-degenerate outer layers Kelvin-Helmholtz contract,
eventually leading to He ignition near their base.  (The M07 model is
similar, though the higher primary WD mass means that He burning
begins almost immediately.)
Once He burning
is ignited off-center there is a $\sim\unit[10^6]{yr}$ phase during
which the He burning migrates to the center through a series of
flashes.  This is in agreement with the results of previous work
\citep{Iben1990, Saio1998}.  The object then enters a long-lived
($\sim\unit[10^8]{yr}$) CHeB phase.  Figure~\ref{fig:gT-and-LT} shows
the surface gravity, effective temperature, and luminosity of our
models during this phase.

After the CHeB phase, there is an off-center He shell-burning phase of
duration $\sim\unit[10^7]{yr}$.  Like the results of
\citet{Zhang2012a}, our \unit[0.5]{\Msun} and \unit[0.7]{\Msun} models
do not experience a cool giant phase (where they would appear like R
Coronae Borealis stars) during shell burning; only a more massive
$\approx \unit[0.8]{\Msun}$ model would have such a phase
\citep{Zhang2012b}.  Once He shell burning ends, the object moves
towards the WD cooling track.  Some of our models experience brief He
flashes before finally moving down the cooling track.  If significant
H survives and is brought to the surface by diffusion, the models have
the H atmospheres of DA WDs; otherwise, they have the He atmospheres
of DB WDs.

\section{Hydrogen}
\label{sec:hydrogen}

We wish to understand the surface H abundance of the WD merger
remnants during the phase when they appear as sdOB stars.  This
observable will be affected by nuclear burning, which decreases the
total amount of H, and by element diffusion, which serves to move the
surviving H towards the surface.  We do not confront the full richness of 
the observational data with our models, instead choosing to initially
focus on the simpler questions of how much H survives and concentrates
at the surface.

Within the hot subdwarf population, there is significant diversity of
the surface H and He abundances.  Most sdB stars have H-rich surfaces
with a population-average He/H number ratio
$\approx 0.01$.\footnote{While low (i.e., sub-solar), such values are
  in fact puzzlingly high. If diffusion were allowed to operate
  unimpeded, the surface layers would become essentially pure H on a
  timescale much shorter than the CHeB lifetime.  This issue and
  possible resolutions, such as mass loss or additional turbulent
  mixing, have been discussed by \citet{Michaud2011} and
  \citet{Hu2011}.}  Asteroseismic analyses of pulsating sdB stars
\citep[e.g.,][]{Brassard2001} allow measurement of the fractional mass
of the outer H-rich envelope; the compilation of \citet{Fontaine2012}
lists measured values for sdBs roughly in the range
$-5.0 \la \log (M_{\rm env}/M) \la -2.5$.  There is also a rare
population of sdB stars with He-dominated surfaces
\citep[He-sdB;][]{Green1986, Ahmad2003, Naslim2010}.  The sdO stars
have both He-deficient and He-enriched sub-classes \citep[e.g.,][]{Hirsch2008}.
Fully successful models of sdOB stars should, for example, reproduce
the correlation between effective temperature and He abundance
\citep{Edelmann2003}.

\subsection{Destruction via H burning}

Hydrogen is not included in the SPH simulations of the double WD
mergers that we use \citep{Dan2011}, but in reality low mass He WDs
have thick H envelopes.  Since we cannot include H self-consistently,
we instead make the simple choice to adopt an initial condition with a
uniform distribution of H in the outer layers (see
Section~\ref{sec:models}).  Physically, this corresponds to the
assumption that turbulent mixing redistributes all the H within these
regions.

\citet{Hall2016} use stellar evolution calculations to characterize
the mass of H expected in the envelopes of the two He WDs at merger.
Across the range of He WD masses and ages they find the H layer mass
is typically 
\unit[$3\times10^{-4}-3\times10^{-3}$]{\Msun}.  Modelling the merger as
a ``cold'' accretion process in which none of this H is destroyed then
gives an upper limit on the amount of H that can be present in the
merged remnant.  However, one expects that some of that H will be
destroyed by the high temperatures during and after the merger.
\citet{Dan2014} use the results of a suite of SPH simulations to
provide fitting formulae that give the fractional mass in each of the
components (core, envelope, and disc) of the merged remnant.
\citet{Hall2016} assume that the H present at merger is distributed
uniformly through the disc and envelope by turbulent mixing and then
assume complete H destruction in the hot envelope.  Since the disc and
envelope components have comparable masses, the surviving H mass is
typically less than the initial H mass by a factor of $\approx 2$.  As
\citet{Hall2016} discuss, this is still an upper limit to the amount
of surviving H.  Modelling the post-merger pre-CHeB phase, as our
work does, is necessary to address further H destruction.

\begin{figure}
  \centering
  \includegraphics[width=\columnwidth]{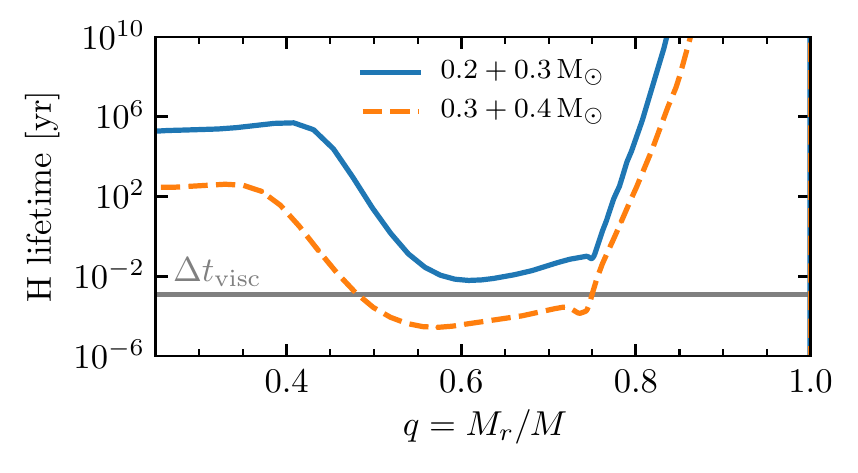}
  \caption{Timescale for H destruction in the initial \MESA\ models.
    The horizontal grey line shows the duration of the viscous phase
    simulations.}
  \label{fig:zp6-tH}
\end{figure}

Figure~\ref{fig:zp6-tH} shows the timescale for H destruction in the
initial conditions of our models.  The length of the viscous
simulation was $\unit[4\times10^4]{s}$ and is marked by the horizontal
grey line.  Around the broad temperature peak ($q \approx 0.6$), H
would be destroyed during the viscous phase or shortly thereafter.
The timescale in the outer layers is much longer; this motivates our
fiducial choice of $\qH = 0.75$.  We also adopt a fiducial abundance
$\XH = 0.01$ as this gives total H masses $\sim\unit[10^{-3}]{\Msun}$,
of order the expected amount \citep[see e.g., figure 12 in][]{Istrate2016b}.

\begin{figure}
  \centering
  \includegraphics[width=\columnwidth]{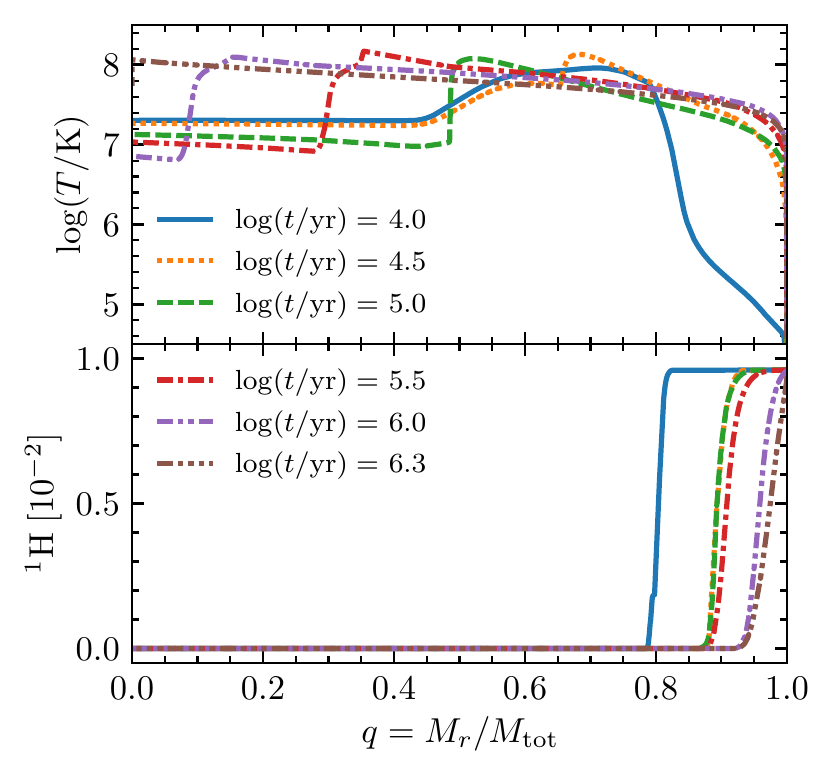}
  \caption{Temperature (upper panel) and H mass fraction (lower panel)
    profiles at the indicated times in model M05.  This model does not
    include the effects of element diffusion.}
  \label{fig:ZP6-TX}
\end{figure}

The upper panel of Figure~\ref{fig:ZP6-TX} shows the temperature
profile of the fiducial model M05 at a sequence of times during its
evolution towards the CHeB phase.  The inward-moving He-burning front
is apparent in the left of the plot.  The structure of the outer
layers evolves such that only a small amount of mass is at
temperatures $\la \unit[10^7]{K}$.  The lower panel shows the profile
of the H mass fraction; most H internal to $q \approx 0.95$ is
destroyed.

\begin{figure}
  \centering
  \includegraphics[width=\columnwidth]{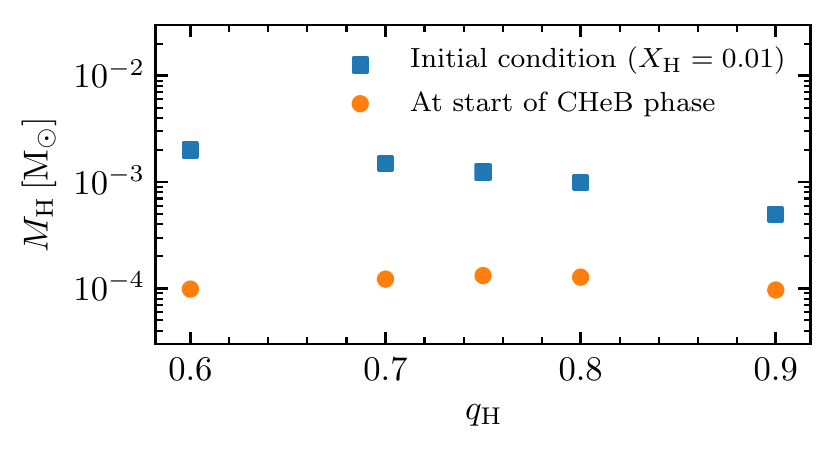}
  \caption{Mass of H surviving at the time model M05 reaches the CHeB
    phase.  The parameter on the x-axis, $\qH$, indicates different
    choices for the initial extent of the H.}
  \label{fig:qH-MH}
\end{figure}

It is a robust feature of our models that H is destroyed throughout a
significant fraction of the material that was originally part of the
disc component.  As such, the surviving hydrogen masses in our models
are much less than those in \citet{Hall2016}.  In contrast to their
models in which approximately half of the H survives post-merger, our
models typically have only roughly a tenth of the initial H remaining.
Figure~\ref{fig:qH-MH} shows the mass of surviving H.  Since H is
always destroyed within the inner regions $(q \la 0.95)$, the mass of
surviving H displays little dependence on its initial distribution.
In all cases shown in Figure~\ref{fig:qH-MH} the surviving H mass is $\approx \unit[10^{-4}]{\Msun}$.
The
more massive model M07 also has surviving H masses
$\approx \unit[10^{-4}]{\Msun}$.
With the assumption of uniform mixing, the absolute mass of surviving
H necessarily depends on the initial mass, since the H present in the outermost layers will survive.  In our models, at fixed $\qH$, a higher value of $\XH$ leads to a larger surviving H mass.
However, for our fiducial value of $\qH = 0.75$, varying the initial value
of $\XH$ by a factor of two in each direction left the percentage of the initial H that survived essentially unchanged at $\approx 10$ per cent.  

\begin{figure}
  \centering
  \includegraphics[width=\columnwidth]{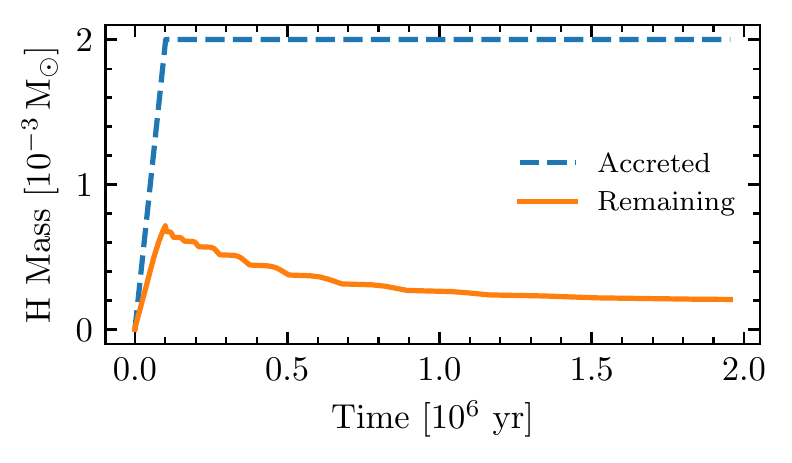}
  \caption{Evolution of an ``accretion'' model similar to our fiducial
    model (see text).  The dashed line shows the cumulative accreted
    mass of H; the solid line shows the surviving H mass.  The
    calculation stops when the model reaches the CHeB phase.}
  \label{fig:acc-mod}
\end{figure}

While H destruction comes naturally from our modelling of the
post-merger phase, we emphasize that this finding is not unique to our
approach.  This same qualitative result can be reproduced in
simulations that model the merger as an accretion event.  To
demonstrate this, we begin with a \unit[0.3]{\Msun} He WD (without a
pre-existing H envelope) and accrete material with $\XH = 0.01$ at a
rate $\unit[2\times10^{-6}]{\Msunyr}$ until the object reaches a mass
of \unit[0.5]{\Msun}.  The object ignites off-center He burning, but
the final total mass is reached before the He burning reaches the
center.  As newly-accreted material is compressed, it reaches
temperatures where the H burns; subsequently, after accretion ceases,
H destruction continues as additional time elapses and the object's
structure adjusts.  Figure~\ref{fig:acc-mod} shows the mass of H that
survives as a function of time.  Most of the accreted H is destroyed
and only $\approx\unit[2\times10^{-4}]{\Msun}$ survives when the model
reaches the CHeB phase.

\subsection{Relocation via Element Diffusion}

In the previous subsection, we showed that H was destroyed throughout
most of the remnant except in the outer layers.  Element diffusion
concentrates the lighter H (in its predominantly He background) at the
surface.  In principle, this could increase the amount of H that
survives the merger by concentrating it in the cooler surface layers
before it has time to burn. However, we find that the action of
element diffusion is not sufficiently rapid to redistribute the H and
prevent its destruction.  Moreover, we find that the action of
rotation (via mixing and mass loss) overwhelms the effects of element
diffusion during the evolution towards the CHeB phase.

To isolate the roles of element diffusion and rotational mixing, we
run a suite of models with and without each of these effects.
Table~\ref{tab:H} summarizes our models and indicates the total
surviving H mass and the H surface abundance in each case (measured at
the start of the CHeB phase).  Models with the same rotation treatment
but with and without element diffusion have nearly identical surviving
H masses.  This is the simple statement that the H diffusion timescale
is not shorter than the H destruction timescale in the deeper layers.
Given the minor effect of rotation and diffusion, we run only one
version of model M07 which uses our fiducial assumptions.

Comparing models with and without rotational mixing, one sees
rotational mixing causes additional H destruction, as this can mix H
into deeper layers.  Allowing element diffusion to operate without the
interference of rotational mixing leads to a slightly higher surface H
fraction.  However, only the completely non-rotating model develops a
surface H abundance $\ga 0.1$.  This reflects the fact that rotating
models typically shed their outer layers (see
Section~\ref{sec:rotation}), meaning that any surface abundance
enhancement does not accumulate.

\begin{table}
\begin{tabular}{lllrr}
Model & Rotation & Diffusion & $M_{\rm H}\;[10^{-4}\,\Msun]$ & $X_{\rm H, surf}$\\
\hline
\multirow{6}{*}{M05}&yes & no & 1.32 & 0.01\\
&yes & yes & 1.31 & 0.01\\
&no mixing & no & 1.96 & 0.01\\
&no mixing & yes & 1.96 & 0.03\\
&no & no & 2.02 & 0.01\\
&no & yes & 1.99 & 0.51\\
\hline
\multirow{1}{*}{M07}&yes & no & 0.96 & 0.01\\
\hline

\end{tabular}
\caption{The effects of rotation and element diffusion.  The column
  ``Rotation'' indicates whether the \MESA\ models include rotation:
  ``yes'' means rotation is included as normal, ``no mixing'' means
  the structural effects of rotation are included, but the rotational
  mixing coefficients are set to 0, and ``no'' means non-rotating models.
  The column ``Diffusion'' indicates whether element diffusion is
  included.  The column $M_{\rm H}$ is the total surviving H mass; the
  column $X_{\rm H, surf}$ is the surface H mass fraction.  Both these
  latter quantities are given at the start of the CHeB
  phase.\label{tab:H}}
\end{table}

\begin{figure}
  \centering
  \includegraphics[width=\columnwidth]{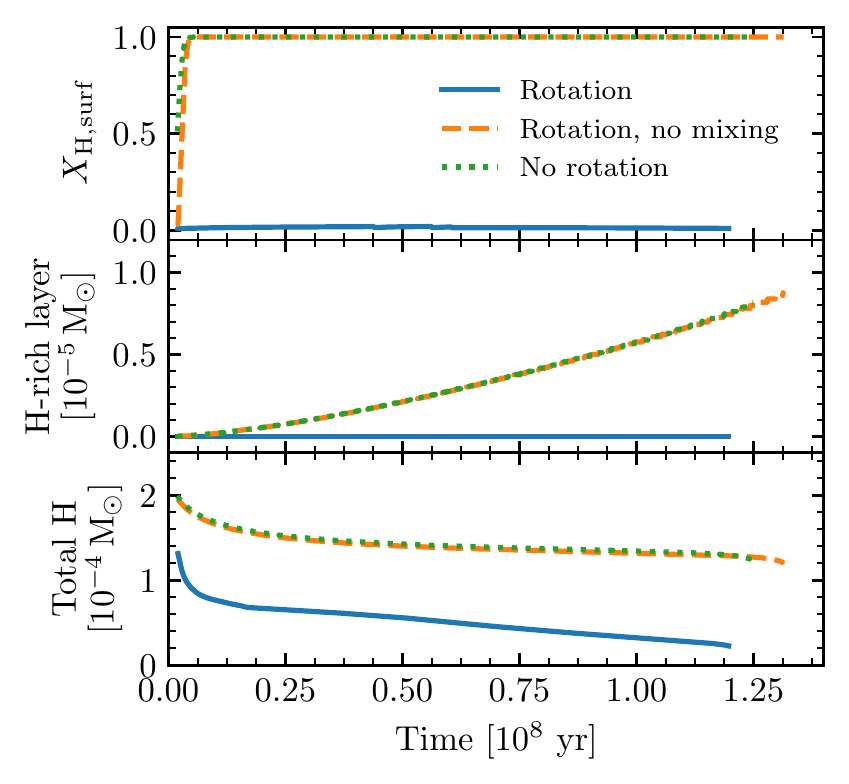}
  \caption{Surface H mass fraction (upper panel), mass in H-rich layer
    (middle panel), and total H mass (lower panel) during
    the CHeB phase.  The
    different lines show runs of model M05 with different treatments
    of rotation and its effects; all models have element diffusion.}
  \label{fig:CHeB-H}
\end{figure}

During the longer lived, higher surface gravity CHeB phase the
diffusion timescale becomes short compared to the evolutionary
timescale.  Figure~\ref{fig:CHeB-H} shows the surface H abundance, the
mass of the H-rich layer (defined as where the H mass fraction
$> 0.1$), and the total H mass for our \MESA\ models during the CHeB
phase.  These are continuations of the three M05 models with
diffusion from Table~\ref{tab:H}.  In cases without rotational mixing,
the surface quickly becomes pure H and the H-rich layer mass grows.
In the case with rotational mixing, this process overwhelms diffusion
and the star does not develop an H-rich surface.  In all cases, some H
continues to be destroyed during the CHeB phase.

The H-rich layer masses that develop when diffusion operates unimpeded
are $\la \unit[10^{-5}]{\Msun}$, even at the end of the CHeB phase.
In the compilation of \citet{Fontaine2012}, several apparently-single
sdB stars have asteroseismically measured envelope masses in excess of
this value.  The work of \citet{Hall2016} also required H masses of
$\sim \unit[10^{-3}]{\Msun}$ to match the atmospheric properties of of
the observed H-rich sdBs.  It is a challenge for our merger models to
explain such objects.

\section{Rotation}
\label{sec:rotation}

The previous section briefly discussed rotation
as a cause of additional mixing.  Here, we wish to understand the
rotation rate of the WD merger remnants during the phase when they
appear as sdOB stars.  The merged system initially retains much of the
orbital angular momentum of the binary.  As discussed by
\citet{Gourgouliatos2006}, if the evolution were strictly
conservative, that amount of angular momentum would cause the star to be
rotating in excess of critical rotation when it reached the CHeB
phase.  During the viscous evolution, angular momentum is transported
outwards, but in the calculations of \citet{Schwab2012} significant
mass and angular momentum were not lost from the system.  Thus, in our
initial conditions, the remnant still has a total angular momentum
that exceeds that of a critically-rotating CHeB star of the same mass.

\begin{figure}
  \centering
  \includegraphics[width=\columnwidth]{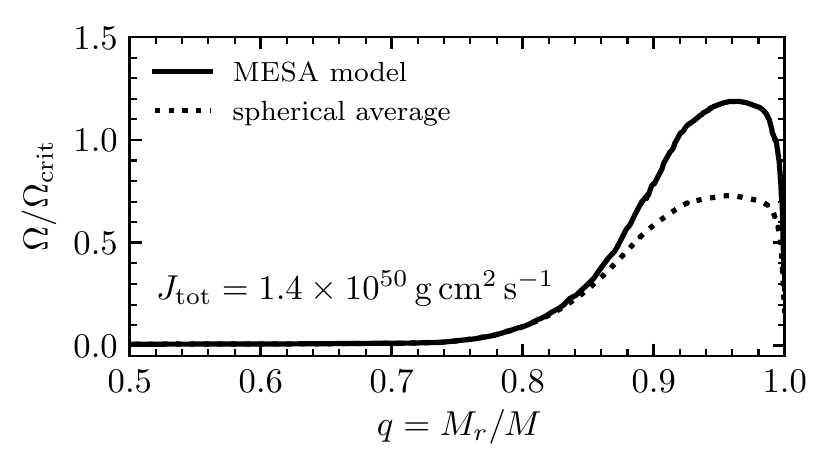}
  \caption{Rotation profile (as a fraction of critical rotation) for
    the initial model M05.  Roughly the outer 10 per cent of material
    is initially super-critical.}
  \label{fig:initial-rotation}
\end{figure}

Figure~\ref{fig:initial-rotation} shows the initial rotation profile
(as a fraction of critical rotation) for our fiducial model along with
the same quantity from the spherical average of the viscous
simulation.  These quantities are not the same, reflecting differences
in the structure of the outer layers that occur when mapping from
multi-D to 1D (see Figure~\ref{fig:zp6-1D}).  More sophisticated
averaging, such as averaging along isobars instead of spherical
shells, might somewhat improve the level of agreement, but we do not
explore this here.  The quantity of primary interest, the total
angular momentum, is the same to within less than 1 per cent.

The fact that material can be above critical rotation in the \MESA\
model reflects the fact that the rotational corrections to the
equations \MESA\ is solving are limited (see discussion in
Section~\ref{sec:models}).  Numerically, this allows for material in
\MESA\ to approach or exceed critical rotation while remaining bound;
physically, we know the behaviour of this material is being poorly
modelled.  This caveat means that the detailed behaviour of the outer
layers during the early phase of evolution is unlikely to be accurate.
However, we argue that the overall picture that develops is robust
against these uncertainties.

These super-critically rotating outer layers serve as an initial
angular momentum reservoir.  The remnant soon expands as it adjusts
towards thermal equilibrium, and since the radius has increased
significantly, this material quickly becomes sub-critical.  Even if
the configuration of the outer layers were different than in our
initial \MESA\ model, we would still expect it to eventually realize a
similar configuration. For example, if some of the high angular
momentum material were instead in a rotationally-supported disc, it
would be quickly engulfed as the star expands and that angular
momentum incorporated in the outer layers.  None of our arguments rely
on the details of the behaviour during this initial
$\sim \unit[10^3]{yr}$ expansion phase.

After realizing its extended state, the remnant begins to contract as
the outer layers cool.  As this occurs, the outer layers again
approach critical rotation.  We allow this mass to be shed from the
star.  The details of this process are uncertain, so we demonstrate
that three mass loss prescriptions all yield similar results.

First, our default approach is to apply the Reimers mass loss rate,
with a scaling factor of 0.1 \citep{Reimers1975}.  This prescription, which
is intended for red giants, is merely a convenient choice for a mass
loss rate that increases with increasing stellar luminosity and
radius.  Following \citet{Heger2000a}, we allow for
rotationally-enhanced mass loss, scaling the mass loss rate by the
factor
\begin{equation}
  \label{eq:H2000}
  \left(1 - \frac{\Omega_{\rm surf}}{\Omega_{\rm surf, crit}}\right)^{-0.43}~.
\end{equation}
By default, \MESA\ caps this enhancement factor at $10^4$.

Our second approach is to begin from the baseline of an arbitrary constant mass loss rate of
$\unit[10^{-10}]{\Msunyr}$.  Again, we allow for rotational enhancement, so the
realized mass loss rate is not constant.  The baseline value is chosen to be low enough that significant
mass would not be lost over the $\sim \unit[10^6]{yr}$ evolution to
the CHeB phase in the absence of rotational enhancement, but to be high enough so that the rotational enhancement factors realized in \MESA\ lead to meaningful loss.
Compared to our default
choice, this approach means that mass is lost at a lower rate for a
longer period.  The slower mass loss means that material is not lost
immediately upon reaching critical rotation and thus there is time for
angular momentum to be redistributed before the mass is shed.  This
leads to less mass and angular momentum being lost than the other
approaches.

The third approach makes use of built-in \MESA\ capabilities that can, at
each timestep, find the mass loss rate that will keep the star
below critical rotation.  This approach avoids invoking any particular
form for the rotationally-enhanced mass loss (such as
equation~\ref{eq:H2000}).  Since the initial condition is
super-critically rotating, we do not activate this mass loss until
after the first $\unit[10^3]{yr}$ have elapsed.  In this case, we also
apply a small constant mass loss rate floor of
$\unit[10^{-10}]{\Msunyr}$ (without rotational enhancement).

\begin{figure}
  \centering
  \includegraphics[width=\columnwidth]{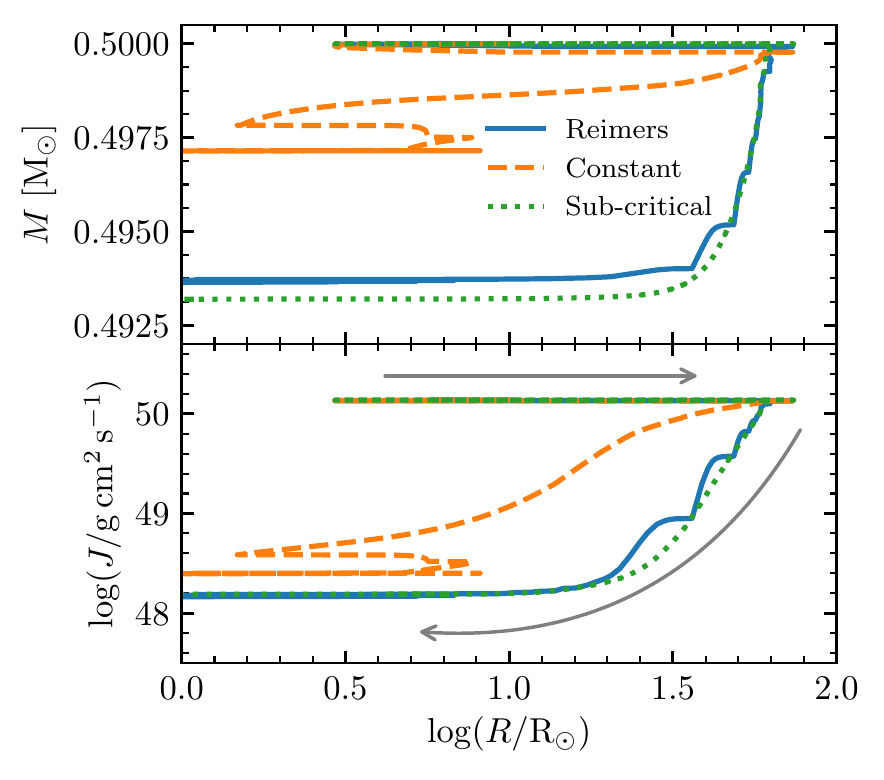}
  \caption{Evolution of the total mass and angular momentum of the
    fiducial model M05 versus its radius.  The grey arrows in the
    lower panel indicate the direction of the evolution.  First, there
    is an expanding phase; this is followed by a contracting phase,
    during which the outer layers reach critical rotation and mass and
    angular momentum are shed.  The three lines correspond to the
    three different mass loss prescriptions described in the text.}
  \label{fig:M-and-J}
\end{figure}

All of the aforementioned prescription are arbitrary and uncertain.
However, the different approaches yield similar results.
Figure~\ref{fig:M-and-J} shows the evolution of the mass and angular
momentum for in calculations using these three prescriptions (labeled
``Reimers'', ``Constant'', and ``Sub-critical'', respectively).  Thus
we argue that so long as the object loses mass in order to remain
at-or-below critical rotation, we expect qualitatively similar
behavior.

As indicated in Figure~\ref{fig:initial-rotation}, the initial total
angular momentum is $\approx \unit[10^{50}]{g\,cm^2\,s^{-1}}$.  If the
shed mass carries the Keplerian angular momentum at the surface, then
the corresponding change in angular momentum is
\begin{equation}
  \label{eq:1}
  \Delta J = \unit[2\times10^{49}]{\rm g\,cm^2\,s^{-1}}
  \left(\frac{R}{\unit[30]{\Rsun}}\right)^{1/2}
  \left(\frac{M}{\unit[0.5]{\Msun}}\right)^{1/2}
  \left(\frac{\Delta M}{\unit[10^{-3}]{\Msun}}\right)~.
\end{equation}
The post-merger expansion of the remnant gives a long lever-arm,
allowing the loss of a small amount of mass,
$\Delta M \sim \unit[5\times10^{-3}]{\Msun}$, to carry away almost all of the
angular momentum.

The results in Figure~\ref{fig:M-and-J} indicate that the amount of
angular momentum retained in the remnant is
$\approx \unit[10^{48}]{g\,cm^2\,s^{-1}}$.  For an
$\approx \unit[0.5]{\Msun}$ CHeB star, the radius is
$\approx \unit[0.1]{\Rsun}$ and moment of inertia is
$\approx \unit[4\times10^{51}]{g\,cm^2}$.  This corresponds to a
surface rotation velocity of
\begin{equation}
  \label{eq:2}
  v_{\rm rot} \approx 20\; {\rm km\,s^{-1}} \left(\frac{J}{10^{48}\;{\rm g\,cm^2\,s^{-1}}} \right)
\end{equation}
Thus, when the \MESA\ models reach the CHeB phase, the remnant is no
longer rapidly rotating, with
$v_{\rm rot} \approx 30\; {\rm km\,s^{-1}}$ for our fiducial model M05.

When we apply the same methods to the model M07, we find that the
initial angular momentum of
$\approx 2 \times \unit[10^{50}]{g\,cm^2\,s^{-1}}$ is reduced to
$\approx \unit[10^{49}]{g\,cm^2\,s^{-1}}$ via the shedding of
$\approx \unit[0.01]{\Msun}$.  Note that the larger amount angular
momentum retained in this case gives a surface rotation velocity
$v_{\rm rot} \approx 100\; {\rm km\,s^{-1}}$.

Observationally, hot subdwarfs---including the apparently-single
ones---are not fast rotators, with most objects having
$v \sin i < \unit[10]{km\,s^{-1}}$ \citep{Geier2012}.  Thus, our
models are still significantly faster rotators than most of the
observed population.  To further reduce the spin, potentially bringing
it in closer accordance with observed values, we would require
additional angular momentum to be shed during post-merger phase or in
the early stages of the CHeB phase.

Angular momentum loss via magnetized stellar winds is a potentially
important process that is not included in our models.  Such winds are
known to be important in the evolution of the rotation rates of main
sequence stars and in the binary evolution of cataclysmic variables
\citep[e.g.,][]{Weber1967, Mestel1968, Verbunt1981}. Their possible
importance in this context was already speculated by
\citet{Iben1986b}.  The magnetic field of the remnant is likely
amplified via dynamo action during the merger \citep{Tout2008, Beloborodov2014}.
Magnetohydrodynamics simulations have found fields in excess of
$\unit[10^8]{G}$ in mergers of two CO WDs \citep{Ji2013, Zhu2015a}.
Determining the field strengths and geometries generated by WD
mergers, including these lower mass He WD systems, will be an
interesting path for future work.

Additionally, \citet{Geier2012} cites an as-yet-unpublished suggestion
by Ph.~Podsiadlowski that angular momentum is shed during the
post-merger He-flashing phase.  In our models, essentially all the
angular momentum loss has already occurred by the time He-flashing
commences, so such a process would represent further loss.

\section{Summary and Conclusions}
\label{sec:conclusions}

We perform stellar evolution calculations of the remnant of the merger
of two He WDs.  We evolve these objects from shortly after the merger
into their core He-burning phase.  At this point the objects appear as
hot subdwarf stars and we follow them until they form WDs.  One of the
key aspects of our work is that we use initial conditions motivated by
hydrodynamic simulations of double WD mergers and the viscous disc
phase that follows.

These initial conditions allow us to capture a phase of expansion
driven by the thermal energy deposited in the remnant during the
merger.  As the energy is radiated away, the star subsequently
contracts.  At that point, the outer layers of the star reach critical
rotation.  We show, via rotating stellar models, that because of the
large stellar radius at this time, a small amount of mass shedding
significantly reduces the total angular momentum of the remnant.  Our
merger model that gives a roughly canonical mass
$(\approx \unit[0.5]{\Msun})$ hot subdwarf is rotating with
$v_{\rm rot} \approx 30\; {\rm km\,s^{-1}}$; the more massive
$\approx \unit[0.7]{\Msun}$ merger remnant is a more rapidly rotating
object with $v_{\rm rot} \approx 100\; {\rm km\,s^{-1}}$.  While these
results demonstrate that the sdOB star remnants of WD mergers need not
be rotating near breakup, they do not yield slowly rotating models in
agreement with most observed objects.  This may indicate the presence
of other efficient angular momentum loss mechanisms not considered in
this work; or it may be evidence that the slowly rotating
apparently-single sdOB stars were not formed through double WD
mergers.  More work is needed to understand whether the trend seen in
these two models implies a mass-rotation relationship, in which higher
mass merger remnants should be faster-rotating.

We also study the ability of H to survive the post-merger evolution.
Calculations of the merger and its aftermath do not include H, so we
adopt a simple initial chemical profile.  We show that if the H is
uniformly distributed in the disc and envelope components of the
merger, it is mostly destroyed.  Beginning with initial H masses of
$\sim \unit[10^{-3}]{\Msun}$ we find surviving H masses
$\approx \unit[10^{-4}]{\Msun}$ for both our \unit[0.5]{\Msun} and
\unit[0.7]{\Msun} models.  This is independent of our assumptions
about rotation and element diffusion.  We also show that simple
accretion models give a similar result.  Moreover, we show that
element diffusion is able to concentrate only a fraction of that H on
the surface over the CHeB lifetime.  At face value, this implies sdOB
stars with H layer masses $\ga \unit[10^{-4}]{\Msun}$ cannot be
formed via WD mergers.  One way around this would be to have the H be
distributed differently than our assumption (i.e., already
concentrated at the surface). Future numerical WD merger calculations
including H may be able to clarify the validity of our assumption
about the initial H distribution.

The initial H layer masses we use are motivated by those found on He
WDs from stellar evolution calculations \citep{Hall2016,
  Istrate2016b}.  Note that during the gravitational-wave-driven inspiral of a He WD
binary, the first mass transferred is necessarily H.  This mass
transfer and accompanying novae may have important consequences on the
pre-merger orbital evolution of the system
\citep[e.g.,][]{Dantona2006, Kaplan2012, Shen2015a}.  Burning of the
transferred mass during this phase can also reduce the amount of H
that is present when the merger occurs, making it even more difficult
to produce remnant with a significant H layer.  These issues further
motivate the careful inclusion of the H envelopes in future work on
these binaries.

In a picture where double He WD mergers produce most single hot
subdwarfs, it is challenging to understand why the the mass and
rotation distributions look similar for sdB stars with and without
observed companions \citep{Geier2012, Fontaine2012}.  Our results do
not alleviate this tension; our models are neither slowly-rotating nor
do they have H envelopes as massive as many of those reported in the
compilation of \citet{Fontaine2012}.  The models of merger-produced
hot subdwarfs in this work bear a greater resemblance to the He-sdO
stars in their rotation rates and abundances \citep[e.g.,][]{Hirsch2009}.

The accretion-based double WD merger models of \citet{Zhang2012a}
successfully reproduce distributions of the surface C and N abundances
observed in the He-sdO stars.  We note that their models generally
have relatively deep convection zones that reach the surface during
the accretion phase (see their figures 13 and 14).  Our models do not
exhibit such convection zones, though our rotating models do have
rotationally-induced mixing processes operating in the outer layers.
This suggests that our models likely have different mixing properties
than theirs, but our parameterized composition profiles limit our
ability to make specific abundance comparisons.  We note that our
\unit[0.5]{\Msun} model has a mass fraction $\la 10^{-3}$ of
\carbon[12] at the surface, while our \unit[0.7]{\Msun} model has
\carbon[12] surface mass fraction $\approx 0.01$.  Both models retain
surface \nitrogen[14] mass fractions around their initial value.
Future work will be necessary to characterize the detailed surface
composition predicted from WD merger models.  Coupled with a better
understanding of the mass-dependence of rotation, future models can
confront observations such as the the finding of \citet{Hirsch2009}
that sdO stars with higher carbon abundances have higher projected
rotational velocities.

\section*{Acknowledgements}

We thank Ken Shen for helpful conversations, encouragement, and
comments on the manuscript.  We thank Marius Dan and Drew Clausen for
useful discussions and acknowledge their involvement during early
phases of this work.  We thank the referee, Zhanwen Han, for his helpful comments.  We thank the participants of the 7th and 8th
meetings on Hot Subdwarfs and Related Objects for stimulating
discussions and Uli Heber for his excellent review articles.  We thank
Pablo Marchant for his work developing the \MESA\ initial model
relaxation routines.  We thank the Los Gatos Public Library, where
much of this work was performed.  This work made use of the Hyades
computing resource supported by NSF AST-1229745.  Support for this
work was provided by NASA through Hubble Fellowship grant \#
HST-HF2-51382.001-A awarded by the Space Telescope Science Institute,
which is operated by the Association of Universities for Research in
Astronomy, Inc., for NASA, under contract NAS5-26555.  This research
has made use of NASA's Astrophysics Data System.  The plotting and analysis was enabled by
\texttt{matplotlib} \citep{hunter_2007_aa},
\texttt{NumPy} \citep{der_walt_2011_aa},
\texttt{ipython/jupyter} \citep{perez2007ipython,kluyver2016jupyter},
\texttt{MesaScript} \citep{MesaScript}, and \texttt{py\_mesa\_reader} \citep{pmr}.


\bibliographystyle{mnras}
\bibliography{wdwd-sdb}


\appendix


\bsp	
\label{lastpage}
\end{document}